\documentclass[12pt]{article}
\usepackage{amsmath}
\usepackage{epsf}
\usepackage{epsfig}
\usepackage{here}
\usepackage{amssymb}
\usepackage{citesort}
\usepackage{graphicx}
\usepackage{latexsym}
\newcommand{\be}{\begin{equation}}
\newcommand{\ee}{\end{equation}}
\newcommand{\bear}{\begin{eqnarray}}
\newcommand{\eear}{\end{eqnarray}}

\newsavebox{\LSIM}
\sbox{\LSIM}{\raisebox{-1ex}{$\ \stackrel{\textstyle<}{\sim}\ $}}
\newcommand{\lsim}{\usebox{\LSIM}}
\newsavebox{\GSIM}
\sbox{\GSIM}{\raisebox{-1ex}{$\ \stackrel{\textstyle>}{\sim}\ $}}
\newcommand{\gsim}{\usebox{\GSIM}}

\begin{document}
\begin{titlepage}
\begin{flushright}
CERN-PH-TH/2005-151\\
hep-ph/0508208
\end{flushright}
$\mbox{ }$
\vspace{.1cm}
\begin{center}
\vspace{.5cm}
{\bf\Large Baryogenesis with Superheavy Squarks}\\[.3cm]
\vspace{1cm}
Stephan J. Huber\footnote{stephan.huber@cern.ch} \\ 
\vspace{1cm}
{\em Theory Division, CERN, CH-1211 Geneva 23, Switzerland} 

\end{center}
\bigskip\noindent
\vspace{1.cm}
\begin{abstract}
We consider a setup where R-parity is violated in the framework of
split supersymmetry. 
The out-of-equilibrium decays of heavy squarks successfully lead to the generation
of a baryon asymmetry.
We restrict the R-parity violating couplings to the baryon number violating
subset to keep the neutralino sufficiently stable to provide the dark matter.
The observed baryon asymmetry can be generated for squark masses
larger than $10^{11}$ GeV, while neutralino dark matter induces a stronger
bound of $10^{13}$ GeV. Some mass splitting between
left- and right-handed squarks may be needed to satisfy also constraints
from gluino cosmology. 
\end{abstract}
\end{titlepage}
\section{Introduction}
The general supersymmetric extension of the standard model (SM) contains baryon and 
lepton number violating interactions \cite{RPV1,RPV2}. Often a discrete
symmetry, called R-parity, is imposed to eliminate these 
operators. In conventional supersymmetric models there are stringent
bounds on many of the R-parity violating couplings induced by
proton decay, neutrino masses, flavor violation, etc. The breaking of
R-parity also has important cosmological implications. The baryon
and lepton number violating interactions may erase a baryon 
asymmetry created in the early universe.  They also induce the
decay of the lightest supersymmetric particle (LSP). Only
in the case of extremely small R-parity violating couplings does the LSP remain
a viable dark matter candidate.

Most of these constraints are considerably relaxed in the context of 
split supersymmetry, where all scalars, except for a single (and finely tuned) 
Higgs boson, have masses much larger than the electroweak scale,
$\tilde m\lsim10^{13}$ GeV \cite{AD04,GR04,ASGR04}. For instance, bounds 
from proton decay have been discussed in ref.~\cite{PFP04}. Taking 
$\tilde m=10^{13}$ GeV, the product of baryon and lepton number violating
couplings can be as large as $10^{-5}$. In conventional supersymmetry
it is bounded to be below $10^{-25}$ to guarantee a sufficiently long-lived
proton. 

The lepton number violating superpotential contains bilinear
terms, $\mu'_iL_iH_2$. They lead to a tree-level mixing between the 
neutrinos and neutralinos which is not suppressed by the heavy mass 
scale of split supersymmetry. This mixing generates a neutrino mass at the tree-level
which constrains the $\mu'$-terms to be smaller than about 1 MeV \cite{CP05}.
It also induces the decay of the lightest neutralino, e.g.~into a neutrino and a (virtual) 
$Z$ boson. Keeping the neutralino lifetime larger than the age of the universe
requires $\mu'_i\lsim10^{-8}$ eV. In contrast, neutralino decay via R-parity 
violating Yukawa couplings (trilinear terms) is indeed suppressed 
by the large sfermion masses \cite{GKM04}. The lepton number  
violating Yukawa couplings induce $\mu'$-terms in the low 
energy effective action by quantum corrections even if they 
are set to zero at the tree-level \cite{BBP95}. Thus,  an enormous
amount of tuning is needed  in this setting to keep the lightest 
neutralino sufficiently stable to provide the dark matter. 

In this article we show how the baryon asymmetry of the universe
could be generated by the out-of-equilibrium decays of heavy 
sfermions in the framework
of split supersymmetry\footnote{See ref.~\cite{alt} for an alternative
proposal.}. In this context, neutralino dark matter is an
important link to tie the masses of the gauginos 
and higgsinos to the electroweak scale \cite{AD04,GR04,ASGR04}. 
To avoid rapid neutralino decay by lepton number violating bilinears,
we restrict R-parity violation to the baryon number violating operators.

There have been attempts to use R-parity violating interactions 
to generate the baryon asymmetry  in weak scale supersymmetry 
\cite{baryold}.
However, these typically require some additional ingredients
to provide the necessary departure from thermal equilibrium,
such as non-thermal production of squarks at the end of inflation,
late decay of gravitinos or axinos, etc. We will show that
in split supersymmetry thermal baryogenesis is successful for squark masses
larger than about $10^{11}$ GeV. As it turns out, the neutralino lifetime
induces a somewhat stronger constraint on the squark masses of
about $10^{13}$ GeV. Gluino cosmology may require  
some mass splitting between
left- and right-handed squarks.
  
The organization of the paper is as follows.
In sec. 2 we discuss how a CP-asymmetry arises in the squark decays. 
It requires the interference of tree-level and two-loop diagrams.
In sec. 3 we write down the Boltzmann equations, which describe the 
baryon production process. Their numerical solution together
with some analytical approximations will be presented in sec. 4. 
Constraints induced by the gluino and neutralino lifetimes will be
discussed in sec. 5. In sec. 6 we present our conclusions.

\section{CP-asymmetry in squark decays}
According to the discussion in the introduction we consider the following 
R-parity violating interactions in the superpotential \cite{RPV1,RPV2}\footnote{Note
that in the literature these couplings are usually denoted by $\lambda_{ijk}''$.}
\be \label{super}
W_{\not\hspace*{-.01cm}R_P}=\lambda_{ijk}U_i^cD_j^cD_k^c,
\ee
where $\lambda_{ijk}=-\lambda_{ikj}$. These operators violate baryon number
but not lepton number. The proton is therefore stable. 
The form of the superpotential (\ref{super}) could, for instance,  be 
guaranteed by lepton parity. 

Since $A$-terms, gaugino masses and the $\mu$-parameter are small 
compared to the split supersymmetry scale $\tilde m$ \cite{AD04,GR04,ASGR04}, scalar 
masses for $U_i^c$ and $D_i^c$ are 
the only soft terms relevant to our discussion. We can diagonalize 
them by supersymmetric rotations. Then flavor transitions can only be
induced by the R-parity violating couplings (\ref{super}) or by ordinary
Yukawa couplings. We make the (conservative) assumption that the effects
of the latter are small, and ignore them.

We assume a (somewhat) hierarchical spectrum of the SU(2) singlet ("right-handed") 
squarks\footnote{For the SU(2) singlet states
we use the notation
$\tilde u\equiv (\tilde u^{c})^{*}$ etc., so that they are counted with 
positive baryon number.}. Then 
baryogenesis is dominated by the decay of the lightest of these states, which 
we take to be the right-handed up squark $\tilde u$. It will turn out that this choice 
simplifies the Boltzmann equations, which describe the baryogenesis process. Taking
another right-handed squark would change the final baryon 
asymmetry only by a factor of order 1.

The spectrum of the SU(2) doublet ("left-handed") squarks is not directly related to 
the baryogenesis
process. If some of them are lighter than $\tilde u$, they only enter the Boltzmann 
equations as additional degrees of freedom, which
carry part of the baryon number. 
Later on this possibility will turn out to be helpful to reconcile the constraints
induced by the gluino and neutralino lifetimes.
A somewhat 
lighter left-handed squark can speed up the gluino decay through supergauge 
interactions, making it more easy to satisfy constraints from gluino cosmology 
\cite{ADGPW05}. The neutralino decay by R-parity violating couplings is still governed 
by the (heavier) right-handed squarks. Thus the neutralino can remain sufficiently 
long-lived to provide the dark matter of the universe.

At least four of the nine couplings of the superpotential (\ref{super}) have 
to be present to allow for 
CP violation in the up squark decays. In the following we assume two of them to be 
$\lambda_{112}$ and $\lambda_{123}$. Two other couplings have to be related to
a different, heavier squark, which we take to be $\tilde t$. The first two operators 
induce baryon number 
violating decays of the up squark. At tree-level the partial decay widths are
\begin{eqnarray}
\Gamma(\tilde u\rightarrow \bar d_1 \bar d_2)\equiv B_{12}\Gamma_D&=&
\frac{|\lambda_{112}|^2}{8\pi}m_{\tilde u}  \nonumber \\
\Gamma(\tilde u\rightarrow \bar d_2 \bar d_3)\equiv B_{23}\Gamma_D&=&
\frac{|\lambda_{123}|^2}{8\pi}m_{\tilde u}  ,
\end{eqnarray}
where $\Gamma_D$ denotes the total decay width and $m_{\tilde u}$ the up squark mass. 
Note that the $d$ quarks are also right-handed 
(SU(2) singlet) states. For simplicity we ignore a possible decay into the
$d_1d_3$ channel.
The up squark also decays by the supergauge interaction at a rate
\be
\Gamma(\tilde u\rightarrow  u \tilde g)\equiv B_g\Gamma_D=\frac{2}{3}\alpha_sm_{\tilde u}  .
\ee
Finally, the up squark can decay through its Yukawa coupling $y_u$.
This contribution is suppressed by the small value of $y_u$ and ignored
in the following. 



At the loop-level, the decay widths for $\tilde u$ and and its antiparticle $\tilde u^*$ become
different in the presence of CP-violation
\begin{eqnarray}
\Gamma(\tilde u\rightarrow \bar d_1 \bar d_2)&=&
\Gamma_D\left(B_{12}-\frac{\epsilon_{12}}{2}\right) 
\nonumber \\
\Gamma(\tilde u\rightarrow \bar d_2 \bar d_3)&=&
\Gamma_D\left(B_{23}-\frac{\epsilon_{23}}{2}\right) 
\nonumber \\
\Gamma(\tilde u\rightarrow \bar   u \tilde g)&=&\Gamma_D\left(B_g+\frac{\epsilon_g}{2}\right).
\end{eqnarray}
Going to antiparticles, the sign of the CP-violating contributions are reversed. 
Since $\tilde u$ and its anti-particle $\tilde u^*$ have the same width, 
\be
\epsilon_g=\epsilon_{12}+\epsilon_{23}
\ee
holds.

\begin{figure}[t] 
\begin{picture}(100,170)
\put(20,20){\epsfxsize6cm \epsffile{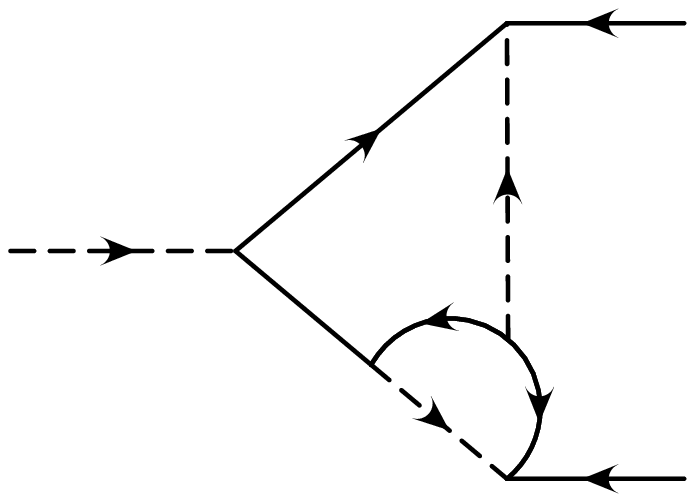}}
\put(240,20){\epsfxsize6cm \epsffile{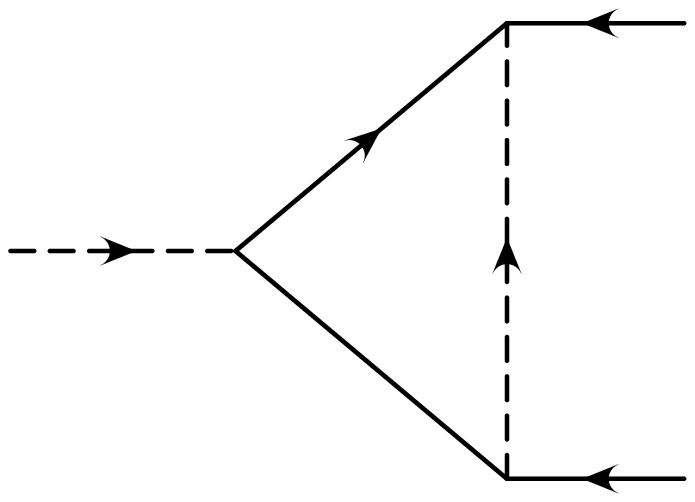}}
\put(30,90){{$\tilde u$}}
\put(250,90){{$\tilde u$}}
\put(177,120){{$\bar d_2$}}
\put(177,30){{$\bar d_1$}}
\put(395,120){{$\bar d_2$}}
\put(395,30){{$\bar d_1$}}
\put(100,115){{$ u$}}
\put(320,115){{$ u$}}
\put(90,50){{$\tilde g$}}
\put(320,40){{$\tilde g$}}
\put(155,90){{$\tilde d_3$}}
\put(375,75){{$\tilde d_3$}}
\put(120,20){{$\tilde t$}}
\put(130,50){{$t$}}
\put(158,50){{$d_2$}}
\put(25,150){{(a)}}
\put(235,150){{(b)}}
\end{picture} 
\caption{2-loop contribution (a) to $\epsilon_{12}$ and its approximation
by a 1-loop diagram (b).}
\label{f_1}
\end{figure}

A CP-asymmetry in the decay width is generated by the interference of tree-level 
and 2-loop amplitudes\footnote{Note that 1-loop contributions are suppressed by the
small gaugino masses and $A$-parameters.}. The diagram of fig.~\ref{f_1}a leads to 
\begin{equation} \label{phase}
\epsilon_{12}\propto{\rm Im}(\lambda_{112}\lambda_{123}^*
\lambda_{312}^*\lambda_{323})\propto \sin(\delta),
\end{equation}
where $\lambda_{312}$ and $\lambda_{323}$ are related to the (heavier) top squark,
and $\delta$ defines the effective CP-phase.
Instead of computing the 2-loop diagram, we approximate the effect of the stop by
a non-diagonal gluino vertex
\begin{equation}
\sqrt{2}\theta_{13}g_3\tilde g \tilde d_1^*d_3.
\end{equation}
Computing the remaining 1-loop diagram of fig.~\ref{f_1}b , we obtain for the CP-asymmetry
\begin{equation} \label{epsilon}
\epsilon_{12}=\frac{2}{3}\alpha_s\sin(\delta)B_{12}\theta_{13}\left| \frac{\lambda_{123}}
{\lambda_{112}}\right|
f\left(\frac{m^2_{\tilde u}}{m^2_{\tilde d_3}}\right).
\end{equation}
The loop function is given by
\begin{equation}
f(x)=2\left(1-\frac{1}{x}\ln(1+x)\right)\approx x+{\cal O}(x^2).
\end{equation}
Of course, there are more diagrams that contribute\footnote{The off-diagonal
gluino vertex can also be generated by ordinary Yukawa couplings,
with left-handed squarks running in the loop.}, so this only a
rough estimate. Note that $\epsilon_{12}$ can be enhanced by taking
the ratio $\lambda_{123}/\lambda_{112}$ to be large.
The CP-asymmetry $\epsilon_{23}$ in the $d_2d_3$ channel is obtained
by $\delta\rightarrow -\delta$ and $m_{\tilde d_3}\rightarrow m_{\tilde d_1}$.


\section{Boltzmann equations}
The evolution of the baryon asymmetry is governed by a set of Boltzmann equations.
We assume that kinetic equilibrium is maintained and approximate the phase-space 
densities by Maxwell-Boltzmann distributions,
\be
f_i(E)=e^{-(E-\mu_i)/T},
\ee
where $\mu_i$ is the chemical potential of the $i$th particle species. It describes the 
deviation from chemical equilibrium.

\subsection{Up squark decay}
The number density of right-handed up squarks is governed by the Boltzmann equation 
\be
\frac{dn_{\tilde u}}{dt}+3Hn_{\tilde u}=-D_g-D_{12}-D_{23}.
\ee
Here $H=1.66\sqrt{g_*}T^2/M_P$ denotes the Hubble parameter, with 
$M_P=1.22\times10^{19}$ GeV. 
If all scalars are heavier than $\tilde u$, the number of degrees of freedom in the plasma is 
$g_*=131.25$. For every  left-handed squark doublet, which is
light, $g_*$ increases by 12. 
To first order in the CP-asymmetries, the collision terms for the different 
decay channels are given by
\begin{eqnarray} \label{ID}
D_g&=&\gamma_D\left[n_{\tilde u}\left(B_g+\frac{\epsilon_g}{2}\right)-
n_{\tilde u}^{\rm eq}(1+\xi_u)\left(B_g-\frac{\epsilon_g}{2}\right)\right] \nonumber \\
D_{12}&=&\gamma_D\left[n_{\tilde u}\left(B_{12}-\frac{\epsilon_{12}}{2}\right)-
n_{\tilde u}^{\rm eq}(1-\xi_{ d_1}-\xi_{ d_2})\left(B_{12}+\frac{\epsilon_{12}}{2}\right)\right] \nonumber \\
D_{23}&=&\gamma_D\left[n_{\tilde u}\left(B_{23}-\frac{\epsilon_{23}}{2}\right)-
n_{\tilde u}^{\rm eq}(1-\xi_{ d_2}-\xi_{ d_3})\left(B_{23}+\frac{\epsilon_{23}}{2}\right)\right],
\end{eqnarray}
where $\xi_i=\mu_i/T$. To obtain this result we approximated $\exp(\xi_i)\approx1+\xi_i$
and used $\xi_{\bar i}=-\xi_i$ for the antiquark chemical potentials. The quark chemical potentials
are first  order in the CP-asymmetries, so that to first order we can drop terms like $\xi\epsilon$.
We used energy conservation to eliminate the 
quark distribution functions
and CPT invariance to obtain the rates for the inverse decay processes.  
The equilibrium number density $n_{\tilde u}^{\rm eq}$ is computed from the 
distribution function with vanishing chemical potential and includes three degrees of 
freedom. 
The thermal decay rate averaged with equilibrium distribution function is \cite{KW80}
\be
\gamma_D(z)=\frac{K_1(z)}{K_2(z)}\Gamma_D,
\ee
where $K_1$ and $K_2$ are Bessel functions and 
\be
z=\frac{m_{\tilde u}}{T}.
\ee
The evolution equation for $\tilde u^*$ is obtained by reversing the signs of the CP-violating
contributions and replacing the quark chemical potentials by that of antiquarks, and vice versa.

It is useful to write the number densities of  $\tilde u$  and $\tilde u^*$ as \cite{KW80}
\be
N_{\pm}=n_{\tilde u}\pm n_{\tilde u^*}.
\ee
Up to linear order in the CP-violating asymmetries, we obtain
\be
\frac{dN_+}{dt}+3HN_+=-\gamma_D(N_+-N_+^{\rm eq})
\ee
The equilibrium density $N_+^{\rm eq}=2n_{\tilde u}^{\rm eq}=(3T^3/\pi^2)z^2K_2(z)$ contains
six degrees of freedom. 

We normalize the number densities to the entropy density $s=(2\pi^2g_*T^3/45)$, where $g_*$
is again the number of degrees of freedom in the plasma, such that
\be
Y_i=\frac{N_i}{s}.
\ee  
Normalization by $s$ removes the term proportional
to the Hubble parameter from the Boltzmann equation. It is convenient to transform to
the dimensionless variable $z$, using $d Y/dt=(z/H(T=m_{\tilde u} ))(d Y/dz)$, to arrive at
\be \label{Yp}
\frac{dY_+}{dz}=-\beta K(Y_+-Y_+^{\rm eq}),
\ee
where
\begin{eqnarray}
\beta(z)&=&z\frac{K_1(z)}{K_2(z)}\nonumber\\
K&=&\frac{\Gamma_D}{H(T=m_{\tilde u})}.
\end{eqnarray}
Assuming that the decay rate is dominated by the gluino channel, we have 
$K\approx 1.5\times 10^3~ (10^{13} {\rm ~GeV}/m_{\tilde u} )$, where the 
running strong coupling is $\alpha_s(10^{13}{\rm ~GeV})=0.035$. A large value
of $K$ signals that the up squark decay is close to equilibrium.
Along the same lines we arrive at
\be \label{Ym}
\frac{dY_-}{dz}=-\beta K(Y_--Y_+^{\rm eq}[B_g\xi_u-B_{12}(\xi_{d_1}+\xi_{d_2})-
B_{23}(\xi_{d_2}+\xi_{d_3})]).
\ee
We see that the inverse decays lead to a back reaction of the
quark densities on the evolution of $Y_-$.

\subsection{Evolution equations for the quark densities}
Before computing the evolution of the quark densities we
have to check what other than the R-parity violating interactions are 
relevant at temperatures
around $10^{13}$GeV which we are going to consider.
\newline
i) Strong sphalerons transform left-handed quarks into right-handed ones.
Using the results of ref.~\cite{SS} and taking into account the running
of $\alpha_s$, we find these processes to be in equilibrium for 
$T\lsim 5\times 10^{13}$ GeV (see also ref.~\cite{B03}).
\newline
ii) Weak sphalerons \cite{WS} become fast at temperatures $T\lsim 2\times 10^{12}$ GeV.
\newline
iii) The scattering rate of Yukawa interactions has been estimated as
$\Gamma_y\approx 0.02y^2T$ \cite{HN95}. The top Yukawa coupling 
therefore is in equilibrium for $T\lsim 10^{16}$ GeV. At  $T\lsim1\times 10^{12}$ GeV
also the bottom and tau Yukawa interactions become relevant.

In the following we will assume the strong sphalerons and the top Yukawa
interaction to be in equilibrium. All other Yukawa couplings and the weak sphalerons 
will be neglected. This approximation is valid in the temperature range
$1\times 10^{12}~{\rm GeV}\lsim T\lsim 5\times 10^{13}$ GeV and leaves
the leptons entirely out of the baryon generation process.

The relevant operators are
\begin{eqnarray} \label{op}
&&\tilde g\tilde u^*u,~\tilde ud_1d_2,~\tilde ud_2d_3,~\tilde g \tilde d_3^* d_1,
\nonumber\\
&&\bar tq_3H,
\nonumber\\
&&uctd_1d_2d_3\bar q_1\bar q_1\bar q_2\bar q_2\bar q_3\bar q_3.
\end{eqnarray}
The quark species $c$, $q_1$ and $q_2$ are only produced by strong
sphalerons. Therefore the number densities are related as
\begin{equation} \label{SS}
2N_c=-N_{q_1}=-N_{q_2}\equiv 2N_q.
\end{equation}
Here we use the number densities of particles minus antiparticles, i.e.~$N_c=n_c-n_{\bar c}$etc.
The operators (\ref{op}) conserve four independent U(1) charges, which  imply
the constraints
\begin{eqnarray} \label{U1}
N_t+N_{q_3}+N_q&=&0
\nonumber\\
N_{d_1}-N_{d_2}+N_{d_3}-N_q&=&0
\nonumber\\
N_-+N_{u}-N_{d_2}&=&0
\nonumber\\
N_t+N_H-N_q&=&0.
\end{eqnarray}
Taking the top Yukawa and strong sphaleron interactions to be in equilibrium
forces the corresponding linear combinations of chemical potentials to vanish.
What these relations mean in terms of number densities depends on the number of 
degrees of freedom (d.o.f.) related  to each chemical potential. To be specific, we assume
that one  Higgs doublet is much heavier than $m_{\tilde u}$ and the supergauge
interactions maintain the equilibrium between the Higgs and higgsino densities. The
Higgs chemical potential then represents 12 d.o.f., like $q_3$. As already mentioned,
we can allow for some left-handed squarks to be light, which we take to be
$\tilde q_1$. Again supergauge interactions will equilibrate the $q_1$ and $\tilde q_1$
densities and effectively double the d.o.f. of $q_1$. The equilibrium conditions 
then take the form
\begin{eqnarray} \label{eq}
-2N_t+N_{q_3}+N_H&=&0
\nonumber\\
N_u+N_t+N_{d_1}+N_{d_2}+N_{d_3}-N_{q_3}+\left(3+\frac{2}{k}\right)N_q&=&0,
\end{eqnarray}
where $k=1~(2)$ if $\tilde q_1$ is heavy (light). Using eqs.~(\ref{SS}), 
(\ref{U1}) and (\ref{eq}) we can express all quantities in terms of $N_-$, $N_u$ and $N_{d_1}$ 
as
\begin{eqnarray} \label{relations}
N_{d_2}&=&N_-+N_u
\nonumber\\
N_{d_3}&=&\frac{3k+2}{5k+2}N_-+\frac{2k+2}{5k+2}N_u-N_{d_1}
\nonumber\\
N_q&=&-N_{q_3}=N_H=-\frac{2k}{5k+2}N_--\frac{3k}{5k+2}N_u
\nonumber\\
N_t&=&0.
\end{eqnarray}
We can use these relations to express baryon number as
\begin{eqnarray}
N_B&\equiv&\frac{1}{3} (N_-+N_u+N_c+N_t+N_{d_1}+N_{d_2}+N_{d_3}+N_{q_1}+
N_{q_2}+N_{q_3})
\nonumber\\
&=&\frac{7k+2}{5k+2}N_-+\frac{8k+2}{5k+2}N_u.
\end{eqnarray}
Note that $N_{d_1}$ does not enter here directly. This is an accident of the approximations
we are using. For instance, it would no longer hold if the bottom Yukawa coupling
were taken into account. 

To compute the generated baryon asymmetry we need two evolution equations in
addition to eqs.~(\ref{Yp}), (\ref{Ym}). We take them to be 
\begin{eqnarray} \label{TE}
\frac{d}{dz}(Y_u-Y_c)&=&D_g-S_{12}-S_{23}-T_{12} -T_{23} -S_{12}'-S_{23}'-T_{12}' -T_{23}'  
- {\rm CP ~conj.}
\nonumber \\
\frac{d}{dz}(Y_{d_1}-Y_c)&=&-D_{12}-S_{12}-T_{12}-S_{13}-S_{12}'-T_{13}-T_{12}' - 
{\rm CP ~conj.}
\end{eqnarray}
Subtracting $Y_c$ in these equations removes the strong sphaleron rate from the
right-hand side, which we have taken to be in equilibrium.
The inverse decays $D_{12}$, $D_{23}$ and $D_g$ are given by eq.~(\ref{ID}).
We can use $n_u-n_{\bar u}\approx 2\xi_u n_q^{\rm eq}$, etc., with $n_q^{\rm eq}=3T^3/\pi^2$, 
to replace the chemical potentials by particle densities, and we define
\begin{equation}
A(z)=\frac{n_{\tilde t}^{\rm eq}}{n_{q}^{\rm eq}}=\frac{z^2}{2}K_2(z).
\end{equation}
The scatterings induced by $\tilde u$ exchange are
\begin{eqnarray}
S_{12}-\bar S_{12}&=&(Y_u+Y_{d_1}+Y_{d_2})\gamma_{S_{12}}+Y_+^{\rm eq}\gamma_D
(B_g\epsilon_{12}+B_{12}\epsilon_g)
\nonumber \\
&&-A(z)B_gB_{12}(Y_u+Y_{d_1}+Y_{d_2})
\nonumber \\[.2cm]
S_{23}-\bar S_{23}&=&(Y_u+Y_{d_2}+Y_{d_3})\gamma_{S_{23}}+Y_+^{\rm eq}\gamma_D
(B_g\epsilon_{23}+B_{23}\epsilon_g)
\nonumber \\
&&-A(z)B_gB_{23}(Y_u+Y_{d_2}+Y_{d_3})
\nonumber \\[.2cm]
S_{13}-\bar S_{13}&=&(Y_{d_1}-Y_{d_3})\gamma_{S_{13}}+Y_+^{\rm eq}\gamma_D
(B_{23}\epsilon_{12}-B_{12}\epsilon_{23})-A(z)B_{12}B_{23}(Y_{d_1}-Y_{d_3})
\nonumber \\[.2cm]
T_{12}-\bar T_{12}&=&2(Y_u+Y_{d_1}+Y_{d_2})\gamma_{T_{12}}
\nonumber \\[.2cm]
T_{23}-\bar T_{23}&=&2(Y_u+Y_{d_2}+Y_{d_3})\gamma_{T_{23}}
\nonumber \\[.2cm]
T_{13}-\bar T_{13}&=&2(Y_{d_1}-Y_{d_3})\gamma_{T_{13}}.
\end{eqnarray}
Here $S_{12}$, $S_{23}$ and $S_{13}$  correspond to the processes 
$t\tilde g\leftrightarrow \bar d_1 \bar d_2$, $t\tilde g\leftrightarrow \bar d_2 \bar d_3$ and
$d_1d_2 \leftrightarrow d_2d_3$. Note that for these s-channel processes  
the resonant parts have been subtracted
since they are already included in the Boltzmann equations \cite{KW80}.
$T_{12}$, $T_{23}$ and $T_{13}$ denote $td_{1,2}\leftrightarrow \tilde g \bar d_{2,1}$, 
$td_{2,3}\leftrightarrow \tilde g \bar d_{3,2}$ and 
$d_{1} \bar d_{2,3} \leftrightarrow d_{3,2}\bar d_{2}$.
Also the exchange of $\tilde d_3$ contributes to the above mentioned scatterings
\begin{eqnarray}
S_{12}'-\bar S_{12}'&=&(Y_u+Y_{d_1}+Y_{d_2})\gamma_{S_{12}'}
\nonumber \\[.2cm]
S_{23}'-\bar S_{23}'&=&(Y_u+Y_{d_2}+Y_{d_3})\gamma_{S_{23}'}
\nonumber \\[.2cm]
T_{12}'-\bar T_{12}'&=&2(Y_u+Y_{d_1}+Y_{d_2})\gamma_{T_{23}'}
\nonumber \\[.2cm]
T_{23}'-\bar T_{23}'&=&2(Y_u+Y_{d_2}+Y_{d_3})\gamma_{T_{23}'}.
\end{eqnarray}
In an approximation we add these contributions incoherently,
which is valid when either $\tilde u$ or $\tilde d_3$ exchange dominates.
There are also contributions to the scatterings by exchange of  $\tilde t$,
which we neglect.

For temperatures considerably smaller than the masses of the exchanged squarks
we obtain for the scatterings  
\bear
\gamma_{S_A}+2\gamma_{T_A}&=&\frac{m_{\tilde u}}{H}\frac{352}{3\pi^2}\lambda_{123}^2
\alpha_s\frac{1}{z^4}
(Y_u+Y_{d_2}+Y_{d_3})
\nonumber \\
\gamma_{S_R}+2\gamma_{T_R}&=&\frac{m_{\tilde u}}{H}\frac{352}{3\pi^2}\lambda_{112}^2
\alpha_s\frac{1}{z^4}
(Y_u+Y_{d_1}+Y_{d_2})
\nonumber \\
\gamma_{S_{13}}+2\gamma_{T_{13}}&=&\frac{m_{\tilde u}}{H}\frac{352}{16\pi^3}\lambda_{112}^2
\lambda_{113}^2\frac{1}{z^4}
(Y_{d_1}-Y_{d_3})
\nonumber \\
\gamma_{S_A'}+2\gamma_{T_A'}&=&\frac{m_{\tilde u}}{H}\frac{352}{3\pi^2}\lambda_{123}^2
\alpha_s\frac{1}{z^4}\frac{m_{\tilde u}^4}{m_{\tilde {d_3}}^4}
(Y_u+Y_{d_2}+Y_{d_3})
\nonumber \\
\gamma_{S_R'}+2\gamma_{T_R'}&=&\frac{m_{\tilde u}}{H}\frac{352}{3\pi^2}\theta_{13}^2
\lambda_{123}^2\alpha_s\frac{1}{z^4}\frac{m_{\tilde u}^4}{m_{\tilde {d_3}}^4}
(Y_u+Y_{d_1}+Y_{d_2}).
\eear
Note that $\gamma_{S_A'}+2\gamma_{T_A'}$ is suppressed by a large factor $m_{\tilde u}^4/m_{\tilde {d_3}}^4\sim1/100$ compared to $\gamma_{S_A}+2\gamma_{T_A}$
 and
can be neglected in the following.

Now we can use the relations (\ref{relations}) to eliminate $Y_{d_2}$ and  $Y_{d_3}$
in the inverse decay
and scattering rates to turn eqs.~(\ref{Ym}), (\ref{TE}) into a closed set of 
equations for $Y_-$, $Y_u$ and $Y_{d_1}$. The result is given in eq.~(\ref{BA})
in the appendix.


\section{Analytical and numerical solutions}
To understand some features of the numerical solution of the Boltzmann
equations  (\ref{BA}), let us first discuss some analytical  approximations.

The departure from equilibrium in the evolution of $Y_+$ is of order $1/K$.
Since for $m_{\tilde u}\lsim10^{14}$ GeV we always have $K\gg1$, we can
expand in powers of $1/K$ (see, for instance ref.~\cite{KT})
To leading order we obtain
\bear
Y_+-Y_+^{\rm eq}&\approx&-\frac{1}{\beta K}\frac{dY_+^{\rm eq}}{dz}
\nonumber\\
&=&\frac{1}{Kz}Y_+^{\rm eq}
\nonumber\\
&=&\frac{135}{2\pi^4g_*}\frac{z}{K}K_2(z)\approx\frac{135}{2^{3/2}\pi^{7/2}g_*}
\frac{\sqrt{z}}{K}e^{-z}.
\eear
Non-equilibrium in the CP-violating densities is governed by some effective
value of $K$, which involves the baryon number violating couplings.

Let us start with the somewhat simpler case $B_{12}=B_{23}$. 
Then the down quark number densities combine as $Y_{d_1}+2Y_{d_2}+Y_{d_3}$, 
since $\gamma_{S_{12},T_{12}}=\gamma_{S_{23},T_{23}}$.
(This is not true for
the scatterings induced by $\tilde d_3$, which however are suppressed
by $m_{\tilde u}^4/m_{\tilde {d_3}}^4\sim1/100$ and can be neglected.)
As a result,  we can form a closed set of equations for $Y_-$ and $Y_B$, which is
given in the appendix (\ref{BABR}). In the limit of large $K$, the equation for 
$Y_-$ reduces to
\begin{equation}
Y_-\approx A(z)\left(\frac{5k+2}{8k+2}-\frac{11k+5}{4k+1}B_R\right)Y_B.
\end{equation}
We also have neglected terms higher order  in $A(z)$, which are exponentially
small at late times, i.e.~for $z\gg1$. We end up with a single equation for the 
baryon asymmetry
\bear
\frac{dY_B}{dz}&=& \label{largeK1}
\beta K\left\{\frac{\epsilon_g}{2}(Y_+-Y_+^{\rm eq})-A(z)B_{12}(1-2B_{12})
\frac{11k+5}{4k+1}Y_B \right\}
\nonumber \\
&&-\frac{11k+5}{4k+1}Y_B\frac{m_{\tilde u}}{H}
\frac{352}{3\pi^2}\lambda_{112}^2\alpha_s\frac{1}{z^4}.
\eear
We observe that the effective value of $K$ is reduced
proportionally to $B_{12}$ and depends somewhat on $k$.

For large values of $B_{12}K$ we can now solve for $Y_B$ using the method 
of steepest descent. Assuming that the
inverse decays dominate, the freeze-out value of
$z$ follows from
\be  \label{largeK2}
\frac{e^{z_f}}{z_f^{5/2}}=\frac{\sqrt{\pi}}{2^{3/2}}\frac{11k+5}{4k+1}B_{12}(1-2B_{12})K,
\ee
which leads to a logarithmic dependence of $z_f$ on $K$. The final baryon asymmetry
is then given by
\be \label{largeK2a}
\eta_B=\frac{28}{79}\frac{\epsilon_g}{2}\frac{135}{\pi^{7/2}
g_*z_fB_{12}(1-2B_{12})K}\frac{4k+1}{11k+5}\sqrt{\frac{4z_f}{2z_f-5}},
\ee
where we added the weak sphaleron factor of 28/79. 
Using eq.~(\ref{epsilon}) we observe that the leading
dependence of $\eta_B$ on $B_{12}$ cancels. This means that smaller values
of the baryon number violating couplings reduce by the same amount the 
CP-asymmetry and the washout by inverse decays.
This compensation works until
$B_{12}K\sim1$. Then the washout is no longer important and the baryon asymmetry
goes to zero proportional to $B_{12}$.
For very large values of $B_{12}K$ the washout is dominated by scatterings. 
Then $z_f\propto K^{1/4}$ and the baryon asymmetry is exponentially damped,
$\eta_B\propto e^{-4z_f/3}$.

In the general case of $B_{12}\neq B_{23}$ eq.~(\ref{largeK1}) generalizes to
\be
\frac{dY_B}{dz}= \label{largeK3}
\beta K\left\{\frac{\epsilon_{12}}{2}(Y_+-Y_+^{\rm eq})-\frac{2}{5}A(z)B_{12}(8-7B_{12}-
8B_{23})Y_B \right\},
\ee
where we have neglected scatterings and restricted ourselves to $\epsilon_{23}=0$
and $k=1$. Note that the washout term is proportional to $B_{12}$, i.e.~the
effective value of $K$ is $K_{\rm eff}\sim B_{12}K$. It can be made
small while keeping a large value of $B_{23}$. This allows us to preserve the $d_1$
part of the baryon asymmetry in the presence of a large $\lambda_{123}$, which can
be used to increase the CP asymmetry. A larger value of $B_{23}$ decreases the
washout. If $\epsilon_{23}$ is non-zero, its contribution to the baryon asymmetry
is washed out with $K_{\rm eff}\sim B_{23}K$, i.e.~it is not protected by a 
small value $B_{12}$.



\begin{figure}[t] 
\begin{picture}(100,160)
\put(-10,180){\epsfxsize6cm \epsffile{m113.eps}}
\put(210,180){\epsfxsize6cm \epsffile{m113A.ps}}
\put(205,130){{$\eta_B$}}
\put(195,115){{$[10^{-10}]$}}
\put(210,145){{$\uparrow$}} 
\put(368,10){$\lambda_{112}\longrightarrow$}
\put(146,10){$\lambda_{112}\longrightarrow$}
\put(100,163){{(a)}}
\put(315,163){{(b)}}
\end{picture} 
\caption{(a) The baryon asymmetry in units of $10^{-10}$ as a function of $\lambda_{112}$ 
for $m_{\tilde u} =5\times10^{13}$ GeV,
$B_{12}=B_{23}$, $k=1$, $\theta_{13}=0.1$ and $m_{\tilde d_3}^2/m_{\tilde u}^2=10$. 
The full (dashed) line includes (neglects) scatterings. The horizontal dashed lines
indicate the observed range.
(b)
The baryon asymmetry with (without) a light $\tilde q_1$ in solid (dashed) line.
Scatterings are neglected.
}
\label{f_2}
\end{figure}

In the numerical evaluations of eq.~(\ref{BA})
we take a maximal CP-phase $\sin \delta=1$,
$\theta_{13}=0.1$ and $m_{\tilde d_3}^2/m_{\tilde u}^2=10$. 
In fig.~\ref{f_2}a we show the baryon asymmetry for $B_{12}=B_{23}$
as a function of $\lambda_{112}$. The other parameters are $k=1$ and
$m_{\tilde u} =5\times10^{13}$ GeV. The dashed horizontal lines indicate
the observed value of the baryon asymmetry, $\eta_B=(0.89\pm0.04)\times10^{-10}$
\cite{BAU}.
The dashed curve takes into account only the inverse decays in the collision
terms of eq.~(\ref{BA}). In agreement with eq.~(\ref{largeK2a}), the baryon 
asymmetry stays
almost constant for  $\lambda_{112}\gsim0.02$. For smaller values the washout
becomes negligible and we have $\eta_B \propto B_{12} \propto\lambda_{112}^2$.
The solid line takes into account  also the scatterings that lead to a strong
damping of the baryon asymmetry for  $\lambda_{112}\gsim0.2$. In fig.~\ref{f_2}b
we investigate the influence of a light left-handed squark $\tilde q_1$ on the baryon 
asymmetry, again neglecting scatterings. The difference is quite small since two
effects partially cancel: the light
$\tilde q_1$ ($k=2$) reduces $\eta_B$ by adding d.o.f. to the plasma. It also
reduces the washout, as we observe from eq.~(\ref{largeK1}).
For smaller up squark masses the maximal baryon asymmetry decreases
approximately as $\eta_B\propto m_{\tilde u}$. So up squark masses above a few times
$10^{13}$ GeV are needed to generate the observed baryon asymmetry
in the case $B_{12}=B_{23}$.

\begin{figure}[t] 
\begin{picture}(100,170)
\put(-10,180){\epsfxsize6cm \epsffile{RNA2.ps}}
\put(212,180){\epsfxsize6cm \epsffile{RNA1.ps}}
\put(205,130){{$\eta_B$}}
\put(195,115){{$[10^{-10}]$}}
\put(210,145){{$\uparrow$}} 
\put(368,10){$\lambda_{112}\longrightarrow$}
\put(146,10){$\lambda_{112}\longrightarrow$}
\put(100,163){{(a)}}
\put(315,163){{(b)}}
\end{picture} 
\caption{
(a)
The baryon asymmetry in units of $10^{-10}$ as function of $\lambda_{112}$ 
for $m_{\tilde u} =1\times10^{13}$ GeV, $\lambda_{123}=0.25$,  $k_{q_1}=1$, 
$\theta_{13}=0.1$ and $m_{\tilde d_3}^2/m_{\tilde u}^2=10$. 
The full (dashed) line includes (neglects) scatterings. 
(b)
The baryon asymmetry with (without) a light $\tilde q_1$ in solid (dashed) line.
Scatterings are neglected.
}
\label{f_3}
\end{figure}

\begin{figure}[t] 
\begin{picture}(100,150)
\put(110,150){\epsfxsize6cm \epsffile{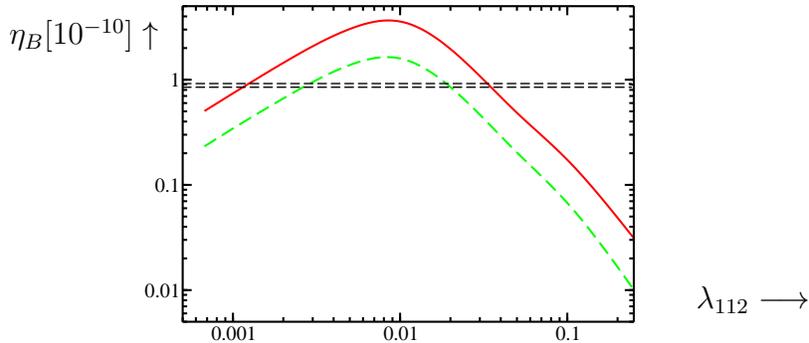}}
\put(70,110){{$\eta_B[10^{-10}]\uparrow$}} 
\put(330,10){$\lambda_{112}\longrightarrow$}
\end{picture} 
\caption{The baryon asymmetry in units of $10^{-10}$ as function of $\lambda_{112}$ 
for $m_{\tilde u} =1\times10^{13}$ GeV,
$k_{q_1}=1$, $\theta_{13}=0.1$ and $m_{\tilde d_3}^2/m_{\tilde u}^2=10$.
Solid (dashed) lines indicate $\lambda_{123}=0.25$ ($0.1$).
Scatterings are included.
}
\label{f_4}
\end{figure}

A larger baryon asymmetry can be produced if $\lambda_{123}>\lambda_{112}$.
In this case $\epsilon_{12}$ is enhanced while the washout remains small due to
$\lambda_{112}$. In fig.~\ref{f_3}a we take $\lambda_{123}=0.25$, 
$m_{\tilde u} =1\times10^{13}$ GeV and $k=1$. Solid (dashed) lines indicate
again that scatterings are included (neglected). Now the baryon asymmetry
rises proportionally to $1/\lambda_{112}$. For $\lambda_{112}\lsim0.01$ the washout
becomes small and $\eta_B\propto \lambda_{112}$. Fig.~\ref{f_3}b shows
that the inclusion of a light $\tilde q_1$ does hardly make any difference. 
In fig.~\ref{f_4} we compare the baryon asymmetry for $\lambda_{123}=0.25$ 
and 0.1. All other parameters are taken as in fig.~\ref{f_3}a. We observe that
$\eta_B$ scales approximately as  $\lambda_{123}$. 

The value of $\lambda_{112}$, where the baryon asymmetry becomes maximal,
i.e.~where the washout becomes ineffective, scales as $m_{\tilde u}^{-1/2}$.
Using this optimal value of $\lambda_{112}$, the baryon asymmetry is therefore 
roughly given by
\be \label{master}
\eta_B\approx10^{-9}\sin(\delta)\lambda_{123}\frac{\theta_{13}}{0.1}
\frac{m_{\tilde u}^2/m_{\tilde d_3}^2}{0.1}
\left(\frac{m_{\tilde u}}{10^{13}{\rm GeV}}\right)^{1/2}.
\ee 
The observed baryon asymmetry can therefore be generated for  
$m_{\tilde u} \gsim10^{11}$ GeV.\footnote{For 
$m_{\tilde u} \lsim10^{12}$ GeV the tau and bottom Yukawa interactions
as well as the weak sphalerons can no longer be neglected. We expect the
corresponding change in $\eta_B$ to be at  most a factor of order unity.} 
For $m_{\tilde u} =1\times10^{13}$ GeV
the coefficient of the off-diagonal gluino vertex can be chosen to be smaller by 
a factor of 10 than used in the figures.

\section{Long-lived neutralinos and gluinos}
By exchange of the lightest right-handed squark, $\tilde u$, the lightest neutralino
predominantly decays as
\begin{equation} \label{decay}
\tilde \chi^0_1\rightarrow ud_2d_3
\end{equation}
and the corresponding antiparticles. Again we assumed that $B_{23}\gg B_{12}$,
so that the decay into $ud_1d_2$ is suppressed. The lifetime associated with this
process is
\begin{equation}
\tau_{\tilde \chi}\approx\frac{1\times10^{16}{\rm yr}}{y^2\lambda_{123}^2}
\left(\frac{m_{\tilde\chi}}{50{\rm GeV}}\right)^{-5}
\left(\frac{m_{\tilde u}}{10^{13}{\rm GeV}}\right)^4.
\end{equation}
where $y$ denotes the coupling $\chi^0_1\bar u \tilde u$. While for a bino-like
$\chi^0_1$ one has $y=\frac{2}{3}\sqrt{2}g_1$, this coupling is given by the up
Yukawa coupling for a higgsino-like LSP, and even vanishes for a pure wino LSP.
Even with $y\lambda_{123}\sim1$ the LSP lifetime is expected to be much
larger than the age of the universe. It can be made even longer lived
by suppressing $y$. If we assume gaugino mass unification, the lightest 
neutralino is predominantly a bino in the range of small $M_2$ \cite{GR04,P04}.
Then we estimate $y^2\sim1/20$.

Stringent constraints on the lifetime of an instable dark matter particle
follow from the production of antiprotons in its decay (\ref{decay}) \cite{BG98}
\begin{equation}
\tau_{\tilde \chi}> 2\times 10^{19}{\rm yr}\left(\frac{m_{\tilde\chi}}{50{\rm GeV}}\right)^{-1}.
\end{equation}
Similar constraints can also be derived from positron production \cite{BJV98}.
This implies that
\begin{equation}
y\lambda_{123}<0.024
\left(\frac{m_{\tilde\chi}}{50{\rm GeV}}\right)^{-2}
\left(\frac{m_{\tilde u}}{10^{13}{\rm GeV}}\right)^2.
\end{equation}
It means that for a bino-like LSP we need $m_{\tilde u}\gsim 10^{13}$ GeV to
be in agreement with the anti-proton bound. Light neutralino masses
are clearly favored. In split supersymmetry there is the possibility of having a 
wino- or higgsino as the LSP, with a mass of 2.0-2.5 and 1.0-1.2 TeV, respectively
\cite{GR04,P04}.
In this case $y$ can be very much suppressed, but it requires some tuning 
to keep the bino content at the level of $10^{-3}$ to compensate for the
larger neutralino masses. It seems to be very difficult to use the 
suppression of $y$ to allow for  $m_{\tilde u}< 10^{13}$ GeV. Moreover,
for very small values of $y$, loop corrections and the contributions from heavier
right-handed squarks become important.

A striking signal of split supersymmetry is the very long-lived gluino.
Its lifetime by R-parity conserving decays has been estimated as \cite{GGS05}
\begin{equation}
\tau_{\tilde g}=4{\rm sec}\times
\left(\frac{m_{\tilde g}}{1{\rm TeV}}\right)^{-5}\times
\left(\frac{\tilde m}{10^{9}{\rm GeV}}\right)^4.
\end{equation}
Heavy isotope searches induce an upper bound on the sfermion mass
scale of $\tilde m\lsim10^{13}$ GeV \cite{AD04}. This is consistent with
the lower bound we previously derived from the neutralino lifetime.
Taking $\tilde m(=m_{\tilde u})=10^{13}$ GeV and $m_{\tilde g}=1$ 
TeV leads to a gluino lifetime 
of about $10^9$ yr. If $B_{23}\sim1$ and $\tilde u$ is the lightest squark, 
the gluino could have a sizable baryon number violating branching ratio. 

Constraints from gluino cosmology depend crucially on the gluino
annihilation cross section after the QCD phase transition. In 
ref.~\cite{ADGPW05} it has been argued that the latter is set by
the de Broglie wavelength of the gluino rather than the geometric
cross section. In this case a stronger bound of   
$\tilde m(=m_{\tilde u})=10^{12}$ GeV can be derived from diffuse gamma rays. 
For $m_{\tilde g}\gsim500$ GeV an even more restrictive bound is induced
by big bang nucleosynthesis. To meet these constraints we would have
to assume a mass splitting between the left- and right-handed squarks.
The right-handed squarks must be heavier than about $10^{13}$ GeV to
keep the neutralino stable enough. Some left-handed squarks, e.g.~$\tilde q_1$
have to be lighter than about  $10^{12}$ GeV to speed up gluino decay.
In this case the baryon number violating branching ratio of the gluino
is highly suppressed by $(\tilde m_{q_1}/\tilde m_{u})^4$. 
Gauge couplings unification is hardly affected by such a mass splitting.
Taking $m_{\tilde q_1}=10^{12}$ GeV, the low energy value of $\alpha_s$ 
is increased by about $2\times10^{-3}$ with respect to the case where
all squarks are degenerate at $\tilde m=10^{13}$ GeV.

\section{Conclusions}
We have investigated baryogenesis by R-parity violating squark decays
in the framework of split supersymmetry. We have restricted ourselves
to the baryon number violating couplings $\lambda_{ijk}$ to avoid rapid 
neutralino decay. These couplings involve only right-handed squarks.
We assume that baryogenesis is dominated by the lightest one, which
we take to be the right-handed up squark.
A CP-asymmetry in the squark decays arises from the interference
of tree-level and 2-loop diagrams involving the CP-phases of $\lambda_{ijk}$.
The relevant Boltzmann equations include, 
in addition to the baryon number violating interactions,
also the supergauge interactions, the strong sphalerons and the top Yukawa interaction.
The generated baryon asymmetry can be enhanced by some hierarchy
in the $\lambda_{ijk}$, where the CP-asymmetry can be increased
while keeping the washout small. 

The observed baryon asymmetry can be successfully generated  if the up squark mass
is larger than $10^{11}$ GeV. A stronger constraint is induced by neutralino decays.
In order to keep the lightest neutralino sufficiently
stable to provide the dark matter, the up squark mass has to be at least  
$10^{13}$ GeV. Depending on the gluino annihilation after the QCD phase
transition, such large squark masses may induce a too large gluino lifetime.
However, somewhat lighter left-handed squarks can speed up the gluino decay.
At the same time the neutralino remains sufficiently stable and baryogenesis
is hardly affected. Favorably,  the LSP is bino-like with some higgsino
admixture and  a mass not far above $M_Z/2$, while the
gluino mass is in the few hundred GeV range. Because of the high sfermion mass
scale, this scenario predicts sizable corrections to chargino and neutralino
Yukawa couplings, which can be probed at a future linear collider \cite{KPRS04}.

It would be interesting to study also the case of the lepton number violating 
couplings. These operators could be responsible for the light neutrino masses
and one might wonder if this is compatible with successful baryogenesis.
However, something would have to be added to the setup to replace the
neutralino as the dark matter particle.

\section*{Appendix}
Using the approximations we discussed in sec.~3, the Boltzmann equations 
that govern the evolution of $Y_-$, $Y_u$ and $Y_{d_1}$ take the form 
\begin{eqnarray}\label{BA}
\frac{d}{dz}Y_-&=&\beta K\left\{-Y_-+A(z)\left[
-\left(B_{12}+\frac{8k+4}{5k+2}B_{23}\right)Y_- \right. \right.
\nonumber \\
&&\left. \left.+\left(1-2B_{12}-\frac{12k+6}{5k+2}B_{23}\right)Y_u
+(B_{23}-B_{12})Y_{d_1}
\right] 
\right\}
\nonumber\\[.8cm] 
\frac{d}{dz}(Y_u-Y_c)&=&\frac{2k}{5k+2}\frac{d}{dz}Y_-+\frac{8k+2}{5k+2}\frac{d}{dz}Y_u=
\nonumber\\[.2cm] 
&=&\beta K\left\{\frac{\epsilon_g}{2}(Y_+-Y_+^{\rm eq})+B_gY_--A(z)B_gY_u \right\}
\nonumber\\
&&-\frac{m_{\tilde u}}{H}\frac{352}{3\pi^2}\lambda_{123}^2
\alpha_s\frac{1}{z^4}\left(1+\frac{m_{\tilde u}^4}{m_{\tilde {d_3}}^4}\right)
\left(\frac{8k+4}{5k+2}Y_-+\frac{12k+6}{5k+2}Y_u-Y_{d_1}\right)
\nonumber\\
&&-\frac{m_{\tilde u}}{H}\frac{352}{3\pi^2}
\alpha_s\frac{1}{z^4}\left(\lambda_{112}^2+\theta_{13}^2\lambda_{123}^2\frac{m_{\tilde u}^4}
{m_{\tilde {d_3}}^4}\right)(Y_-+2Y_u+Y_{d_1})
\nonumber \\[.8cm] 
\frac{d}{dz}(Y_{d_1}-Y_c)&=&\frac{2k}{5k+2}\frac{d}{dz}Y_-+\frac{3k}{5k+2}\frac{d}{dz}Y_u+
\frac{d}{dz}Y_{d_1}=
\nonumber\\[.2cm] 
&=&\beta K\left\{\frac{\epsilon_{12}}{2}(Y_+-Y_+^{\rm eq})-B_{12}Y_--A(z)B_{12}(Y_-+Y_u+
Y_{d_1}) \right\}
\nonumber\\
&&-\frac{m_{\tilde u}}{H}\frac{352}{3\pi^2}
\alpha_s\frac{1}{z^4}\left(\lambda_{112}^2+\theta_{13}^2\lambda_{123}^2\frac{m_{\tilde u}^4}
{m_{\tilde {d_3}}^4}\right)(Y_-+2Y_u+Y_{d_1})
\nonumber\\
&&-\frac{m_{\tilde u}}{H}\frac{352}{16\pi^3}\lambda_{112}^2
\lambda_{113}^2\frac{1}{z^4}
\left(2Y_{d_1}-\frac{3k+2}{5k+2}Y_--\frac{2k+2}{5k+2}Y_u\right).
\end{eqnarray}
The scatterings are given for temperatures considerably smaller than $m_{\tilde u}$, 
i.e.~$z\gg1$.

In the special case $B_{12}=B_{23}$, the down number densities always appear
in the combination $Y_{d_1}+2Y_{d_2}+Y_{d_3}$, since 
$\gamma_{S_{12},T_{12}}=\gamma_{S_{23},T_{23}}$.
(This is not true for
the scatterings induced by $\tilde d_3$, which however are suppressed
by $m_{\tilde u}^4/m_{\tilde {d_3}}^4\sim1/100$ and can be neglected.)
Thus we can form a closed set of equations for $Y_-$ and $Y_B$
\begin{eqnarray} \label{BABR}
\frac{dY_-}{dz}&=&\beta K\left\{-Y_-+A(z)
\left[\left(-\frac{7k+2}{8k+2}+\frac{7k}{4k+1}B_{12}\right)Y_-+\left(\frac{5k+2}{8k+2}
+\frac{11k+5}{4k+1}B_{12}\right)Y_B\right] \right\}
\nonumber \\[.5cm]
\frac{dY_B}{dz}&=&
\beta K\left\{\frac{\epsilon_g}{2}(Y_+-Y_+^{\rm eq})-2B_{12}Y_--A(z)B_{12}
\left[\frac{2k}{4k+1}Y_-+\frac{6k+3}{4k+1}Y_B\right] \right\}
\nonumber \\
&&-\left(-\frac{5k+2}{4k+1}Y_-+\frac{11k+5}{4k+1}Y_B\right)\frac{m_{\tilde u}}{H}
\frac{352}{3\pi^2}\lambda_{112}^2\alpha_s\frac{1}{z^4}.
\end{eqnarray}

\section*{Acknowledgements}
I would like to thank
L.~Covi, A.~Riotto, A.~Ritz, A.~Romanino, Z.~Tavartkiladze, P.~Uwer and O.~Vives
for helpful discussions. Special thanks to G.~Giudice for several
fruitful discussions and collaboration in the early stages of this work.


\begin{thebibliography}{12}
\bibitem{RPV1}
S.~Weinberg, {\em Phys.~Rev.} {\bf D26} (1982) 287; N.~Sakai and T.~Yanagida, 
{\em Nucl.~Phys.} {\bf B197} (1982) 533.
\bibitem{RPV2}
For a recent review, see e.g.~R. Barbier {\em et al.}, hep-ph/0406039.
                 
\bibitem{AD04}
N.~Arkani-Hamed and S.~Dimopoulos, hep-th/0405159. 

\bibitem{GR04}
G.F.~Giudice and A.~Romanino, {\em Nucl.~Phys.} {\bf B699} (2004) 65 [hep-ph/0406088],
{\em Erratum ibid.} {\bf B706} (2005) 65.

\bibitem{ASGR04}
N.~Arkani-Hamed, S.~Dimopoulos, G.F.~Giudice and A.~Romanino,
{\em Nucl.~Phys.} {\bf B709} (2005) 3 [hep-ph/0409232]. 


\bibitem{PFP04}
P.~F.~Perez, {\em J.~Phys.} {\bf G31} (2005) 1025 [hep-ph/0412347].

\bibitem{CP05}
E.J.~Chun and S.C.~Park, {\em JHEP} {\bf 0501} (2005) 009 [hep-ph/0410242].

\bibitem{GKM04}
S.K.~Gupta, P.~Konar and B.~Mukhopadhyaya, {\em Phys.~Lett.} {\bf B606}
(2005) 384 [hep-ph/0408296].

\bibitem{BBP95}
 V.D.~Barger, M.S.~Berger, R.J.N.~Phillips and T.~Wohrmann, {\em Phys.~Rev.}
{\bf D53} (1996) 6407 [hep-ph/9511473];  
B.~de Carlos and P.L.~White, {\em Phys.~Rev.} {\bf D54} (1996) 3427 [hep-ph/9602381].

\bibitem{alt}
S.~Kasuya and F.~Takahashi, {\em Phys.~Rev.} {\bf D71} (2005) 121303 [hep-ph/0501240].

\bibitem{baryold}
S.~Dimopoulos and  L.J.~Hall, {\em Phys.~Lett.} {\bf B196} (1987) 135;
J.~Cline and S.~Raby, {\em Phys.~Rev.} {\bf D43} (1991) 1781;
R.J.~Scherrer, J.~Cline, S.~Raby and D.~Seckel,
{\em Phys.~Rev.} {\bf D44} (1991) 3760;
S.~Mollerach and E.~Roulet, {\em Phys.~Lett.} {\bf B281} (1992) 303;
A.~Masiero and  A.~Riotto, {\em Phys.~Lett.} {\bf B289} (1992) 73 [hep-ph/9206212];
R.~Adhikari and U.~Sarkar, {\em Phys.~Lett.} {\bf B427} (1998) 59 [hep-ph/9610221].


\bibitem{ADGPW05}
A.~Arvanitaki, C.~Davis, P.W.~Graham, A.~Pierce and J.G.~Wacker, hep-ph/0504210.

\bibitem{KW80}
E.W.~Kolb and S.~Wolfram, {\em Nucl.~Phys.} {\bf B172} (1980) 224, 
{\em Erratum-ibid.} {\bf B195} (1982) 542. 

\bibitem{SS}
G.~D.~Moore, {\em Phys.~Lett.} {\bf B412} (1997) 359 [hep-ph/9705248]. 

\bibitem{B03}
L.~Bento, {\em JCAP} {\bf 0311} (2003) 002 [hep-ph/0304263].

\bibitem{WS}
G.D.~Moore, {\em Phys. Rev.} {\bf D62} (2000) 085011 [hep-ph/0001216], hep-ph/0009161.

\bibitem{HN95}
P.~Huet and A.E.~Nelson, {\em Phys.~Rev.} {\bf D53} (1996) 4578 [hep-ph/9506477]. 

\bibitem{KT}
E.W.~Kolb and M.S.~Turner, {\em The Early Universe}, Addison-Wesley Publishing
Company, 1990. 

\bibitem{BAU}
D.N.~Spergel {\em et al.}, WMAP Collaboration, {\em Astrophys.~J.~Suppl.}
{\bf 148} (2003) 175 [astro-ph/0302209];
M.~Tegmark {\em et al.}, SDSS Collaboration,
{\em Phys.~Rev.} {\bf D69} (2004) 103501 [astro-ph/0310723]. 

\bibitem{P04}
A.~Pierce, {\em Phys.~Rev.} {\bf D70} (2004) 075006 [hep-ph/0406144].

\bibitem{BG98}
E.A.~Baltz and P.~Gondolo, {\em Phys.~Rev.}~{\bf D57} (1998) 7601 [hep-ph/9704411]. 

\bibitem{BJV98}
V.~Berezinsky, A.~Masiero and  J.W.F.~Valle, {\em Phys.~Lett.}~{\bf B266}(1991) 382;
V.~Berezinsky, A.S. Joshipura and J.W.F.~Valle, {\em Phys.~Rev.}~{\bf D57} (1998) 147 [hep-ph/9608307]. 


\bibitem{GGS05}
P.~Gambino, G.F.~Giudice and P.~Slavich, hep-ph/0506214.

\bibitem{KPRS04}
W.~Kilian, T.~Plehn, P.~Richardson and E.~Schmidt, {\em Eur.~Phys.~J.} {\bf C39}
(2005) 229 [hep-ph/0408088].
\end{thebibliography}
\end{document}